\documentstyle{article}
\def\nonequal{\,=\kern -0.9em /\,\,}
\begin{document}
{}~\hfill hep-th/9911016
\begin{center}
{\Large \bf Light cone formalism in AdS spacetime}
\footnote{
Talk given at  XIV International Workshop
on High Energy Physics and Quantum Field Theory
QFTHEP'99, Moscow, Russia, 27 May - 2 June, 1999
}
\\

\vspace{4mm}

R.R.Metsaev\\
Department of Theoretical Physics, P.N. Lebedev Physical
Institute, Leninsky prospect 53, 117924, Moscow, Russia
\end{center}

\begin{abstract}
Light cone form of field dynamics in anti-de Sitter spacetime
is described. We also present light cone reformulation of the boundary
conformal field theory representations. AdS/CFT correspondence between the
bulk fields and the boundary operators is discussed.  \end{abstract}


{\bf Motivation}. A long term motivation to develop light cone formalism in
AdS spacetime comes from a number of the following potentially important
applications.

One important application is to type IIB superstring in $AdS_5\times S^5$
background. Inspired by  the conjectured duality between the string theory
and ${\cal N}=4$, $4d$ SYM theory \cite{mal} the Green-Schwarz formulation
of strings propagating in $AdS_5\times S^5$ was suggested in \cite{mt1}
(for  further developments see \cite{krr}-\cite{k1}).  Despite
considerable efforts these strings have not yet been quantized (some
related interesting discussions are in \cite{Dolan:1999pi},\cite{Ber}).
Alternative approaches can be found in \cite{alter}. As is well known,
quantization of GS superstrings propagating in flat space is
straightforward only in the light cone gauge.  The light cone gauge in
string theory implies the corresponding light cone formulation for target
space fields.  In the case of strings in AdS background this suggests that
we should first study a light cone form dynamics of {\it target space
fields} propagating in AdS spacetime. Understanding  a light cone
description of AdS target space fields might help to solve problems of
strings in AdS spacetime.

The second application is to a  theory of higher massless spin fields
propagating in AdS spacetime. Some time ago completely self-consistent
interacting equations of motion for higher massless fields of all spins in
four-dimensional AdS spacetime have been discovered \cite{vas1}. For
generalization to higher spacetime dimensions see \cite{vas2}. Despite
efforts the action that leads to these equations of motion has not yet
been obtained.  In order to quantize these theories and investigate their
ultraviolet behavior it would be important find an appropriate action.
Since the higher massless spin theories correspond quantum mechanically to
non-local point particles in a  space of certain auxiliary variables, it
is conjectured  that they may be ultraviolet finite (see
\cite{vas3},\cite{vas4}). We believe that a light cone formulation is
what is required to understand these theories better.  The situation here
may be analogous to that in string theory; a covariant formulation of
closed string field theories is non-polynomial and is not useful for
practical calculations, while  the light cone  formulation restricts the
string action to cubic order in string fields.

{\bf Light cone formulation of field dynamics in AdS spacetime}.
First let us discuss the forms of AdS algebra, that is $so(d-1,2)$,
we are going to use.  AdS algebra of $d$ dimensional AdS
spacetime consists of translation generators $\hat{P}^A$ and rotation
generators $\hat{J}^{AB}$ which span $so(d-1,1)$ Lorentz algebra. The
commutation relations of AdS algebra in antihermitean basis are
$$
[\hat{P}^A,\hat{P}^B]=\lambda^2\hat{J}^{AB}\,,
\quad
[\hat{P}^A,\hat{J}^{BC}]=\eta^{AB}\hat{P}^C-\eta^{AC}\hat{P}^B\,,
\quad
[\hat{J}^{AB},\hat{J}^{CE}]=\eta^{BC}\hat{J}^{AE}
+3\  \hbox{terms}\,,
$$
where $\eta^{AB}=(-,+\ldots,+)$, $A,B,C,E=0,1,\ldots,d-1$.
The $\lambda$ is a cosmological constant of AdS
spacetime. As $\lambda\rightarrow 0$
the AdS algebra becomes the Poincar\'e algebra
$$
\lim_{\lambda\rightarrow 0} \hat{P}^A=P_{Poin}^A\,,
\qquad
\lim_{\lambda\rightarrow 0} \hat{J}^{AB}=J_{Poin}^{AB}\,.
$$
This form algebra is not convenient for our purposes.
We prefer to use the form provided by nomenclature
of conformal algebra. Namely we introduce new basis
which consists of new translations $P^a$,
conformal boosts $K^a$, dilatation $D$,
and $so(d-2,1)$ algebra generators $J^{ab}$ defined by
$$
P^a\equiv \hat{P}^a+\lambda\hat{J}^{d-2a}\,,
\qquad
K^a\equiv \frac{1}{2}(-\frac{1}{\lambda^2} \hat{P}^a
+\frac{1}{\lambda}\hat{J}^{d-2 a})\,,
\qquad
D\equiv -\frac{1}{\lambda}\hat{P}^{d-2}
\qquad
J^{ab}\equiv \hat{J}^{ab}\,.
$$
Flat space limit in this notation is given by
\begin{equation}\label{contract2}
\lim_{\lambda\rightarrow 0}
P^a=P_{Poin}^a\,,
\qquad
\lim_{\lambda\rightarrow 0}
(-\lambda D)=P_{Poin}^{d-2}\,,
\qquad
\lim_{\lambda\rightarrow 0}
(\frac{1}{2\lambda}P^a +\lambda K^a)
=J_{Poin}^{d-2 a}\,.
\end{equation}
In the conformal algebra basis one has the following
well known commutation relations
\begin{eqnarray}
\label{ppkk}
&
{}[D,P^a]=-P^a\,,
\qquad
[D,K^a]=K^a\,,
\qquad
[P^a,P^b]=0\,,
\qquad
[K^a,K^b]=0\,,
&
\\
&
{}[P^a,J^{bc}]=\eta^{ab}P^c
-\eta^{ac}P^b\,,
\qquad
[K^a,J^{bc}]=\eta^{ab}K^c-\eta^{ac}K^b\,,
\\
\label{pkjj}
&
{}[P^a,K^b]=\eta^{ab}D-J^{ab}\,,
\qquad
[J^{ab},J^{ce}]=\eta^{bc}J^{ae}+3\hbox{ terms}\,,
\qquad
\eta^{ab}=(-,+,\ldots+)\,,
&
\end{eqnarray}
where $a,b,c,e=0,1,\ldots,d-3,d-1$.
In this form the AdS algebra is known as the algebra of
conformal transformations in $(d-1)$-dimensional Minkowski
spacetime. Below we shall be interested in realization of
this algebra as the one of transformations of massless bulk
fields propagating in $d$-dimensional AdS spacetime as well as in
realization of this algebra as the algebra of conformal
transformations on appropriate CFT operators in $(d-1)$-dimensional
Minkowski spacetime.

To develop light cone formulation we shall use
Poincar\'e parametrization of AdS spacetime in which
$$
ds^2=\frac{1}{z^2}(-dt^2+dx_i^2+dz^2+ dx_{d-1}^2)\,,
\qquad
z>0\,.
$$
Here and below we set cosmological constant $\lambda$ equal to unity.
The boundary at spatial infinity corresponds to $z=0$.\footnote{
Poincar\'e coordinates cover half of AdS spacetime. Because we
are interested in infinitesimal transformation laws of physical fields
the global description of AdS spacetime is not important for our study.}
The Killing vectors in these coordinates are given by\footnote{The
target space indices $\mu,\nu$ take the values $0,1,\ldots,d-1$.}
\begin{eqnarray}\label{kilvec}
\xi^{P^a,\mu}=\eta^{a\mu}\,,
\quad
\xi^{K^a,\mu}=-\frac{1}{2}x_\nu^2\eta^{a\mu}+x^a x^\mu\,,
\quad
\xi^{D,\mu}=x^\mu\,,
\quad
\xi^{J^{ab},\mu}
=x^a\eta^{b\mu} - x^b\eta^{a\mu}\,,
\end{eqnarray}
while the corresponding generators are defined as
$G=\xi^{G,\mu}\partial_\mu$. Then we introduce light cone variables
$x^\pm$, $x^I=(x^i,x^z)$
$$
x^\pm \equiv\frac{1}{\sqrt{2}}(x^{d-1}\pm x^0)\,,
\qquad
x^0\equiv t,\,\,\,x^{d-2} \equiv z\,,
\qquad
I,J,K,L=1,\ldots,d-2\,,
\qquad
i,j,k,l=1,\ldots,d-3\,.
$$
In this notation scalar product of tangent space vectors is  decomposed as
$$
X^A Y^A = X^+Y^- + X^-Y^+ +X^IY^I\,,
\qquad
X^IY^I=X^iY^i+X^zY^z\,,
$$
i.e. we use the convention $X^{d-2}=X^z$. The coordinate $x^+$ is
considered as an evolution parameter. Here and below to simplify our
expressions we will drop the metric tensors $\eta_{AB}$, $\eta_{ab}$ in
scalar products.

In light cone basis the AdS algebra splits into generators

\begin{equation}\label{kingen}
P^+,\,P^i,\, J^{+i},\, K^+,\, K^i,\, D,\, J^{+-},\,J^{ij}\,,
\end{equation}
which we refer to as kinematical generators and
\begin{equation}\label{dyngen}
P^-,\, J^{-i}\,, K^-\,,
\end{equation}
which we refer to as dynamical generators. For $x^+=0$ the kinematical
generators are realised quadratically in physical fields while the
dynamical generators receive corrections in interaction theory. In this
paper we deal with free fields.  The light cone form of AdS algebra can be
obtained from (\ref{ppkk})-(\ref{pkjj}) with the light cone metric having
the following non
vanishing elements $\eta^{+-}=\eta^{-+}=1$, $\eta^{ij}=\delta^{ij}$.

Now our primary goal is to find realization of this algebra on
the physical fields. For definiteness we will be interested in
spin $s$ totally symmetric fields.
To keep formulas as simple as possible, let us start with
spin one Maxwell field. Instead of target space field $A^\mu$
with the equations of motion $D_\mu F^{\mu\nu}=0$ we introduce tangent
space field $\Phi^A$ defined by
\begin{equation}\label{tantar}
\Phi^A\equiv e^A_\mu A^\mu,
\qquad
e_\mu^A=\frac{1}{z}\delta_\mu^A
\end{equation}
and use equations of motion in tangent space
$D_BF^{BA}=0$, $F_{AB}=D_A \Phi_B - D_B\Phi_A$,
where $F^{AB}$ is the field strength in the tangent space while
$D_A$ is covariant derivative which in Poincar\'e coordinates take the form
$$
D_A\Phi_B=\hat{\partial}_A\Phi_B
+\delta_B^z\Phi_A-\eta_{AB}\Phi_z\,,
\qquad\qquad
F_{AB}=\hat{\partial}_A \Phi_B-\hat{\partial}_B\Phi_A
+\delta_B^z\Phi_A-\delta_A^z\Phi_B\,,
\quad
\hat{\partial}_A\equiv e_A^\mu\partial_\mu
$$
From these relations one gets the following second order equations of
motion for the gauge field $\Phi^A$
\begin{equation}\label{eqmot3}
(\hat{\partial}^2+(1-d)\hat{\partial}_z+d-2)\Phi^A
-\hat{\partial}^A(\hat{\partial}\Phi)
+(d-3)\hat{\partial}^A \Phi^z
+(2-d)\delta_z^A \Phi^z+2\delta_z^A(\hat{\partial}\Phi)=0\,,
\end{equation}
where $\hat{\partial}^2\equiv \hat{\partial}_A^2$ and
$\hat{\partial}\Phi\equiv \hat{\partial}^A\Phi^A$.  Since these equations
are invariant with respect to the gauge transformation $\delta \Phi^A
=\hat{\partial}^A\Lambda$ we can impose the light cone gauge
\begin{equation}\label{lcv}
\Phi^+=0\,.
\end{equation}
Inserting this into equations (\ref{eqmot3}) we get the following
constraint\footnote{Recall that in the Minkowski spacetime the Maxwell
equations in gauge $\Phi^+=0$ lead to the Lorentz constraint $\partial^A
\Phi^A=0$.  This is not the case in AdS spacetime. Here, by virtue of the
relation $D_A\Phi^A=\hat{\partial}\Phi+(1-d)\Phi^z$, the constraint
(\ref{gcon}) does not coincide with the Lorentz constraint $D^A\Phi^A=0$.}
\begin{equation}\label{gcon}
\hat{\partial}^A\Phi^A=(d-3)\Phi^z\,.
\end{equation} From (\ref{gcon}) we express the $\Phi^-$
in terms of the physical field\footnote{We assume, as usual in light cone
formalism, that the  operator $\partial^+$ has trivial kernel.}
\begin{equation}\label{Phmi}
\Phi^- = -\frac{\partial^I}{\partial^+}\Phi^I
+\frac{d-3}{\hat{\partial}^+}\Phi^z\,.
\end{equation}
Note that the second term in r.h.s. of equation (\ref{Phmi}) is absent
in flat space.  It is this term that breaks $so(d-2)$ manifest invariance
and reduce it to $so(d-3)$ one. By virtue of the constraint
(\ref{gcon}) the equations of motion (\ref{eqmot3}) take the form
$$
(\hat{\partial}^2+(1-d)\hat{\partial}_z+d-2)\Phi^A
+(d-4)\delta_z^A \Phi^z=0\,.
$$ From this we get the following equations for the physical fields
$\Phi^i$, $\Phi^z$:
$$
(\hat{\partial}^2+(1-d)\hat{\partial}_z+d-2)\Phi^i=0\,,
\qquad
(\hat{\partial}^2+(1-d)\hat{\partial}_z+2d-6)\Phi^z=0\,.
$$
Since this form of equations of motion is not convenient we introduce
new physical field $\phi^I$ defined by\footnote{Note that it is the field
$\phi^I$ that has conventional canonical dimension $\Delta_0=(d-2)/2$.}
\begin{equation}\label{Phphv}
\Phi^I=z^{(d-2)/2}\phi^I\,.
\end{equation}
In terms of $\phi^I$ the equations of motion take the form
\begin{equation}
\label{eqmotphii}
(\partial^2-\frac{1}{4z^2}(d-2)(d-4))\phi^i=0\,,
\qquad
(\partial^2-\frac{1}{4z^2}(d-4)(d-6))\phi^z=0\,.
\end{equation}
Dividing by $\partial^+$ these equations can be rewritten in the
Schrodinger form
\begin{equation}\label{schequ}
\partial^-\phi^I =P^-\phi^I\,,
\end{equation}
where the action of $P^-$ on physical fields is defined by
\begin{equation}
\label{hamvec1}
P^-\phi^i=\Bigl(-\frac{\partial_I^2}{2\partial^+}
+\frac{(d-2)(d-4)}{8z^2\partial^+}\Bigr)\phi^i\,,
\qquad
P^-\phi^z=\Bigl(-\frac{\partial_I^2}{2\partial^+}
+\frac{(d-4)(d-6)}{8z^2\partial^+}\Bigr)\phi^z\,.
\end{equation}
Few comments are in order. i) From equations of motion (\ref{eqmotphii})
we see that in $d=4$ the mass like terms cancel.  This fact reflects the
conformal invariance of spin one field in four dimensional AdS spacetime;
ii) In the equations above the field $\phi^i$ and $\phi^z$ have different
mass term. This is reflection of breaking the usual flat space transverse
$so(d-2)$ algebra. On the other hand we see that AdS light cone
equations of motion keep manifest $so(d-3)$ symmetry.
Thus light cone formalism in AdS spacetime breaks manifest $so(d-2)$
symmetry and keeps manifest $so(d-3)$ one.

Before continuing with our main theme
let us note that gauge invariant action for spin one
Maxwell field $S=-\frac{1}{4}\int\! d^dx\sqrt{g}F_{AB}^2$
takes the following simple form in terms of physical field
$\phi^I$
\begin{equation}\label{lcac} S_{l.c.}=\int d^dx
\partial^+\phi^I(-\partial^-+P^-)\phi^I\,.
\end{equation}

Now let us turn to our primary aim that is transformation of physical
degrees of freedom $\phi^I$. Toward this aim we start, as usual, with
original global  AdS symmetries, supplemented with
compensating gauge transformations required to maintain the gauge
$$
\delta_{tot} A^\mu
=\xi^\nu\partial_\nu A^\mu
-A^\nu\partial_\nu \xi^\mu
+\partial_\mu \Lambda\,.
$$
As usual the gauge transformation parameter $\Lambda$ can be found from
the requirement $\delta_{tot}\Phi^+=0$.
From these relations transformation of physical field $\phi^I$ is fixed
uniquely. The expressions can be simplified if we
 use creation and annihilation
operators $\alpha^I$, $\bar{\alpha}^I$,
$[\bar{\alpha}^I,\alpha^J]=\delta^{IJ}$,
and introduce Fock vector for the physical field
$|\phi\rangle\equiv \phi^I\alpha^I|0\rangle$.
The transformation laws of the physical field can be then cast into the
following form

\begin{equation}\label{fintr}
\delta_{tot}|\phi\rangle
=\Bigl(\xi\partial+\frac{d-2}{2z}\xi^z
+\frac{1}{2}\partial^I\xi^JM^{IJ}
+M^{IJ}\partial^+\xi^I\frac{\partial^J}{\partial^+}
-\frac{\partial^+\xi^I}{2z\partial^+}\{M^{zj},M^{jI}\}
\Bigr)|\phi\rangle
\end{equation}
where the spin operator $M^{IJ}$ is given by
\begin{equation}\label{rab}
M^{IJ}\equiv \alpha^I\bar{\alpha}^J-\alpha^J\bar{\alpha}^I\,.
\end{equation}
Generalization to the case of arbitrary spin $s$ totally symmetric field
is relatively straightforward (see \cite{metlc}). In this case the
physical degrees of freedom are described by traceless totally symmetric
tensor field $\phi^{I_1\ldots I_s}$. To simplify expressions it is useful
as above to collect physical degrees of freedom in one Fock vector
$$
|\phi\rangle\equiv
\phi^{I_1\ldots I_s}\alpha^{I_1}\ldots \alpha^{I_s}|0\rangle\,,
\qquad
\bar{\alpha}_I^2|\phi\rangle=0\,.
$$
With this notation the equations of motion and corresponding
hamiltonian $P^-$ take the form
\begin{equation}\label{symham}
\Bigl(z^2\partial^2+\frac{1}{2}M_{ij}^2
-\frac{(d-4)(d-6)}{4}\Bigr)|\phi\rangle
=0,\qquad
P^-=-\frac{\partial_I^2}{2\partial^+}
+\frac{1}{2z^2\partial^+}(-\frac{1}{2}M_{ij}^2+\frac{(d-4)(d-6)}{4})\,.
\end{equation}
The transformations of physical fields under the action of global AdS
symmetries take the same form as in (\ref{fintr}) where we have to use
Schrodinger equation of motion (\ref{schequ}) and $P^-$ given in
(\ref{symham}).\footnote{Note that light cone action for arbitrary spin
$s$ field takes the following elegant and extremely simple form\\
$S_{l.c.}=\int d^dx \langle\partial^+\phi|
(-\partial^-+P^-)|\phi\rangle$ where $P^-$ is given by
(\ref{symham}).
In $d=4$ because of $M_{11}=0$ the mass like term in (\ref{symham})
cancels. This fact reflects conformal invariance of arbitrary spin
$s$ totally symmetric massless field in $AdS_4$ spacetime.}
Making use of the AdS transformations (\ref{fintr})
we can represent them as differential operators acting on the
physical massless field $|\phi\rangle$. Plugging the Killing vectors
(\ref{kilvec}) in transformation laws (\ref{fintr}) we get
corresponding differential form of generators.  Light cone form of AdS
algebra kinematical generators is given by
\begin{eqnarray} \label{3spi} &&{}\hspace{-1cm}
P^i=\partial^i\,,
\qquad
P^+=\partial^+\,,
\qquad
D=x^+ P^-+x^-\partial^++x^I\partial^I+\frac{d-2}{2}\,,
\\
\label{3sjpm}
&&{}\hspace{-1cm}
J^{+-}=x^+P^--x^-\partial^+\,,
\qquad
J^{+i}=x^+\partial^i-x^i\partial^+\,,
\qquad
J^{ij}=x^i\partial^j-x^j\partial^i+M^{ij}\,,
\\
\label{3skp}
&&{}\hspace{-1cm}
K^+=-\frac{1}{2}(2x^+x^-+x_I^2)\partial^++x^+ D\,,
\qquad
K^i=-\frac{1}{2}(2x^+x^-+x_J^2)\partial^i+x^i D+M^{iI}x^I+M^{i-}x^+\,,
\end{eqnarray}
where
\begin{equation}\label{mmi}
M^{-i}=M^{ij}\frac{\partial^j}{\partial^+}
-\frac{1}{\partial^+}M^i\,,
\quad
M^i\equiv M^{zi}\partial_z+\frac{1}{2z}\{M^{zj},M^{ji}\}.
\end{equation}
$P^-$ is given in (\ref{symham}) and remaining dynamical generators are
given by
\begin{eqnarray}
\label{adsjmi}
&&
J^{-i}
=x^-\partial^i-x^i P^- +M^{-i}\,,
\\
\label{3skm}
&&
K^-=-\frac{1}{2}(2x^+ x^-+x_I^2) P^- +x^- D
+ \frac{1}{\partial^+}x^I\partial^J M^{IJ}
-\frac{x^I}{2z\partial^+}\{M^{zJ},M^{JI}\}\,.
\end{eqnarray}

{\bf Light cone form of conformal field theory}.
Now the $so(d-1,2)$ algebra is considered as
algebra of conformal transformations of $(d-1)$-dimensional Minkowski
spacetime and we are interested in light cone reformulation of (free)
conformal field theory. The reason for doing  this is that we are going to
establish AdS/CFT correspondence between bulk massless fields and
conformal field theory operators. As the bulk massless fields have been
studied within the framework of the light cone formalism the most
adequate form for comparison is the light cone form of conformal field
theory. To keep our presentation as simple as possible we restrict our
attention to the case of arbitrary spin totally symmetric operators
${\cal O}^{a_1\ldots a_s}(x)$ that have canonical conformal dimension
$\Delta=s+d-3$.\footnote{The fact that this expression is
nothing but the lowest energy value of spin $s$ massless fields
propagating in $d$ dimensional AdS spacetime has been demonstrated in
\cite{metsit1}.} These operators, by definition, are traceless
${\cal O}^{aaa_3\ldots a_s}=0$ and divergence free
$\partial_{x^a}{\cal O}^{aa_2\ldots a_s}=0$.
As above to simplify our presentation we consider Fock space vector
(generating function)
$|{\cal O}_{cov}\rangle
\equiv {\cal O}^{a_1\ldots a_s}\alpha^{a_1}\ldots\alpha^{a_s}|0\rangle$.
In terms of generating function the traceless and divergence free
conditions take the following form
\begin{equation}
\label{cfttra1}
\bar{\alpha}^a\bar{\alpha}^a|{\cal O}_{cov}\rangle=0\,,
\qquad
\bar{\alpha}^a\partial_{x^a}|{\cal O}_{cov}\rangle=0\,.
\end{equation}
Realization of conformal algebra generators on the space of
operators $|{\cal O}_{cov}\rangle$ is given by \cite{macksal}
\begin{eqnarray}
\label{covpa}
&&
P^a= \partial^a\,,
\qquad
K^a=-\frac{1}{2}x_b^2\partial^a+x^a(x^b\partial_{x^b}+\Delta)
+M^{ab}x^b\,,
\\
\label{covjab}
&&
J^{ab}=x^a \partial^b-x^b\partial^a+M^{ab}\,,
\qquad
D=x^a\partial_{x^a}+\Delta\,,
\qquad
M^{ab}=\alpha^a\bar{\alpha}^b-\alpha^b\bar{\alpha}^a\,,
\end{eqnarray}
where $\partial^a\equiv\eta^{ab}\partial_{x^b}$ and
 $M^{ab}$ is the $so(d-2,1)$ algebra spin operator.
Because in the bulk the $so(d-1,2)$ algebra was realized on the space of
unconstrained physical fields it is reasonable to solve the second
constraint in (\ref{cfttra1}) and formulate boundary conformal theory also
in terms of unconstrained operators which we shall denote by $|{\cal
O}\rangle$.  One can choose a basis which makes $J^{+i}$, $J^{+-}$ and
$K^+$independent of $M^{ab}$.  The  final light cone form of the
generators realized on conformal theory operators is (see \cite{metlc})
\begin{eqnarray}
\label{cftpa}
&&
P^a=\partial^a\,,
\qquad
D=x^+\partial^- + x^-\partial^+ + x^i\partial^i
+\hat{\Delta}\,,
\\
\label{cftjpi}
&&
J^{+i}=x^+\partial^i-x^i\partial^+\,,
\qquad
J^{+-}=x^+\partial^--x^-\partial^+\,,
\qquad
J^{ij}=x^i\partial^j - x^j\partial^i+M^{ij}\,,
\\
\label{cftkp}
&&
K^+=-\frac{1}{2}(2x^+x^- + x_i^2)\partial^+
+x^+D\,,
\qquad
J^{-i}=x^-\partial^i-x^i\partial^-
+M^{ij}\frac{\partial^j}{\partial^+}
-\frac{1}{\partial^+}M^i\,,
\end{eqnarray}
where the spin operator $M^{ij}$ is the same as in (\ref{covjab}).
The operator $M^i$ transforms in vector representation of the spin
operator $M^{ij}$ and satisfies the commutation relations\footnote{Detailed
description of operator $M^i$ which should not be confused with $M^i$
given in (\ref{mmi}) may be found in \cite{metlc}.}
$$
[M^i,M^{jk}]=\delta^{ij}M^k-\delta^{ik}M^j\,,
\qquad
[M^i,M^i]=\Box M^{ij}\,,
$$
where $\Box$ is the Dalamber operator in $(d-1)$ dimensional
Minkowski spacetime $\Box\equiv \partial_{x^a}^2$.
A representation of spin part $\hat{\Delta}$ of the
dilatation operator  $D$ on conformal
operator $|{\cal O}\rangle$ is determined to be
\begin{equation}\label{cftdsp}
\hat{\Delta}\equiv\alpha^i\bar{\alpha}^i+d-3\,.
\end{equation}

{\bf AdS/CFT correspondence}.
After we have derived the light cone formulation for both the bulk
field $|\phi\rangle$ and the boundary conformal theory operator
$|{\cal O}\rangle$ we are ready to demonstrate
explicitly the AdS/CFT correspondence.  We demonstrate that boundary
values of normalizable solutions of bulk equations of motion are
related to conformal operators $|{\cal O}\rangle$
(see \cite{Balasubramanian:1999sn},\cite{metlc}).

To this end we decompose field $|\phi\rangle$ into irreducible
representations of $so(d-3)$ subalgebra
$|\phi\rangle=\sum_{s^\prime=0}^s \oplus |\phi_{s^\prime}\rangle$,
$(\alpha^i\bar{\alpha}^i-s^\prime)|\phi_{s^\prime}\rangle
=\bar{\alpha}_i^2|\phi_{s^\prime}\rangle=0$,
and rewrite equations of motion (\ref{symham}) in the form
\begin{equation}\label{equmot}
(-\partial_z^2+\frac{1}{z^2}(\kappa^2-\frac{1}{4}))|\phi_{s^\prime}\rangle
=q^2 |\phi_{s^\prime}\rangle\,,
\qquad
\kappa\equiv s^\prime+\frac{d-5}{2},
\qquad
q\equiv \sqrt{\Box}\,,
\end{equation}
where we have used the relation
$$
\frac{1}{2}M_{ij}^2|\phi_{s^\prime}\rangle
=-s^\prime(s^\prime+d-5)|\phi_{s^\prime}\rangle
$$
Normalizable solution to the equation (\ref{equmot}) is
\begin{equation}\label{solequmot}
|\phi_{s^\prime}(x,z)\rangle
=\sqrt{qz}J_\kappa (qz)
q^{-(\kappa+\frac{1}{2})}|{\cal O}_{s^\prime}(x)\rangle
\end{equation}
where an operator $|{\cal O}\rangle$\footnote{Note
that we decompose the operator $|{\cal O}\rangle$ into
irreducible representations of $so(d-3)$ algebra
$|{\cal O}\rangle=\sum_{s^\prime=0}^s |{\cal O}_{s^\prime}\rangle$,
where $|{\cal O}_{s^\prime}\rangle$ satisfy the relations
$(\alpha^i\bar{\alpha}^i-s^\prime)|{\cal O}_{s^\prime}\rangle
=\bar{\alpha}_i^2|{\cal O}_{s^\prime}\rangle=0$.}
does not depend on $z$ and the $J_\kappa$ is Bessel
function. In
(\ref{solequmot}) we use the notation $|{\cal O}\rangle$ since we are
going to demonstrate that this is indeed the CFT operator discussed above.
Namely,  we are going to prove that AdS transformations for $|\phi\rangle$
lead to conformal theory transformations for $|{\cal O}\rangle$
\begin{equation}\label{gadsgcft} G_{ads}|\phi_{s^\prime}\rangle =
Z_\kappa(qz)q^{-(\kappa+\frac{1}{2})}G_{cft}|{\cal O}_{s^\prime}\rangle\,,
\qquad Z_\kappa(z)\equiv \sqrt{z}J_\kappa(z)\,.
\end{equation}
Here and below we use the notation $G_{ads}$ and $G_{cft}$ to indicate the
realization of $so(d-1,2)$ algebra generators on the  bulk field
(\ref{3spi})-(\ref{3skm}) and conformal operator
(\ref{cftpa})-(\ref{cftdsp}) respectively.

Toward this purpose let us make a comparison for
generators for bulk field $|\phi\rangle$ and
boundary operator $|{\cal O}\rangle$. Important technical simplification is
that it is sufficient to make comparison only for the part of the algebra
spanned by generators
$P^a$, $J^{ab}$, $D$, $K^+$.
It is straightforward to see that if these generators match then the
remaining generators $K^-$ and $K^i$
shall match due to commutation relations of the $so(d-1,2)$
algebra.  We start with a comparison of the kinematical generators
(\ref{kingen}).  As for  the generators $P^+$, $P^i$, $J^{+i}$,
$J^{ij}$, they already
coincide on both sides (see (\ref{3spi}),(\ref{3sjpm}) and
(\ref{cftpa}),(\ref{cftjpi})). In fact this implies that the
coordinates $p^+$, $p^i$, we use on AdS and CFT sides match.

Now let us consider $P_{ads}^-$ (\ref{symham}) and $P_{cft}^-$
(\ref{cftpa}). Taking into account that the solutions to equations
of motion (\ref{symham}), by definition, satisfy the relation
$P_{ads}^-|\phi\rangle=\partial^-|\phi\rangle$ we get
that $P_{ads}^-$ and $P_{cft}^-$ satisfy the desired relation
(\ref{gadsgcft}). So $P^-_{ads}$ and $P_{cft}^-$ also match.  Taking this
into account it is straightforward to see that the generators
$J_{ads}^{+-}$, $D_{ads}$ (\ref{3sjpm}),(\ref{3spi}) and $J^{+-}_{cft}$,
$D_{cft}$ (\ref{cftjpi}), (\ref{cftpa}) also satisfy the relation
(\ref{gadsgcft}). Next we consider the kinematical generators $K^+_{ads}$
(\ref{3skp}) and $K_{cft}^+$ (\ref{cftkp}).  Making use of the
following relation
$$
K_{ads}^+
q^{-(\kappa+\frac{1}{2})}Z_\kappa(qz)
=q^{-(\kappa+\frac{1}{2})}Z_\kappa(qz)
(K_{cft}^+ +x^+z\partial_z)
-q^{-(\kappa+\frac{1}{2})}\frac{\partial^+}{q}
(\partial_qZ_\kappa(qz))z\partial_z\,,
$$
the relation (\ref{solequmot}) and the fact that
$|{\cal O}\rangle$ in (\ref{solequmot}) does not depend $z$ we get
immediately that $K_{ads}^+$ and $K_{cft}^+$ satisfy the relation
(\ref{gadsgcft}).
The last step is to match the generators
$J_{ads}^{-i}$ and $J_{cft}^{-i}$.
Using the relation $P_{ads}^-=P_{cft}^-$ and comparing the above
expression for $J_{ads}^{-i}$ with $J_{cft}^{-i}$ given in
(\ref{adsjmi}) and (\ref{cftkp}) we conclude that all that remains to do
is to match $M_{ads}^i$ (\ref{mmi}) and $M_{cft}^i$.
Technically, this is the most
difficult point of matching which can be proved by direct calculation (see
\cite{metlc}).

Finally let us write $AdS/CFT$ correspondence for bulk symmetric spin
$s$ fields and corresponding boundary conformal theory operators.
From (\ref{solequmot}) we can read, up to a factor, the following relation
$$
\lim_{z\rightarrow 0} z^{-\hat{\Delta}+\Delta_0}
|\phi_{s^\prime}(x,z)\rangle
=|{\cal O}_{s^\prime}(x)\rangle
$$
where in this expression $\hat{\Delta}$ is the spin part of dilatation
generator given in (\ref{cftdsp})
while $\Delta_0$ is a canonical dimension of bulk massless field in
$AdS_d$ spacetime $\Delta_0=(d-2)/2$.
Above analysis is straightforwardly generalized to the case of
non-normalizable solutions and shadow operators (see \cite{metlc}).

{\bf Conclusions} We have developed  the light cone formalism in AdS
spacetime.  Here we discussed application of this formalism to the  study
of AdS/CFT correspondence but it is applicable for discussion of various
interesting problems some of which are:
(i) generalization to supersymmetry and applications to
supergravity in $AdS$ background and then to strings in this background;
(ii) extension of light cone formulation of conformal field
theory to the level of OPE's and study of light cone form of AdS/CFT
correspondence at the level of correlation functions;
 (iii) application of
light cone formalism to the study of the S-matrix along the lines of
\cite{polch}--\cite{gidd};
(iv) applications to interaction vertices for
higher massless spin fields in AdS spacetime.
Because the formalism we presented
is algebraic in nature it allows us to treat  fields with arbitrary spin
on equal footing.  Comparison of this formalism with other approaches
available in the literature leads us to  the conclusion that this is a
very efficient formalism. Application of the formalism above to study
of $IIB$ supergravity in $AdS_5\times S^5$ background may be found in
\cite{metiib}.

{\bf Acknowledgments}. This  work was supported in part
by INTAS grant No.96-538 and the Russian Foundation for Basic Research
Grant No.99-02-17916.


\end{document}